\documentclass[prl,twocolumn,superscriptaddress,showpacs,unsortedaddress]{revtex4}

\usepackage{graphicx}% Include figure files
\usepackage{dcolumn}% Align table columns on decimal point
\usepackage{bm}% bold math
\usepackage{epsfig}
\usepackage{graphicx}
\usepackage{textcomp}
\usepackage{amsmath}
\usepackage{amssymb}
\usepackage{subfigure}

\begin{document}

%\title{Magnetic structure refinement using circularly polarized X-rays with application to multiferroic TbMnO$_3$}
\title{Circularly polarised X-rays as a probe of non-collinear magnetic order in multiferroic TbMnO$_3$}

\author{F. Fabrizi}
\affiliation{ European Synchrotron Radiation Facility, B\^{o}ite Postale 220, 38043 Grenoble, France }
\affiliation{London Centre for Nanotechnology and Department of Physics and Astronomy, University College London, Gower Street, London WC1E 6BT, UK}

\author{H. C. Walker}
 \email{helen.walker@ucl.ac.uk}
\affiliation{London Centre for Nanotechnology and Department of
Physics and Astronomy, University College London, Gower Street,
London WC1E 6BT, UK}

\author{L. Paolasini}
\affiliation{ European Synchrotron Radiation Facility, B\^{o}ite
Postale 220, 38043 Grenoble, France }

\author{F. de Bergevin}
\affiliation{ European Synchrotron Radiation Facility, B\^{o}ite Postale 220, 38043 Grenoble, France }

\author{A. T. Boothroyd}
\affiliation{ Department  of Physics, Clarendon Laboratory, University of Oxford, UK}

\author{D. Prabhakaran}
\affiliation{ Department  of Physics, Clarendon Laboratory, University of Oxford, UK}

\author{D. F. McMorrow}
\affiliation{London Centre for Nanotechnology and Department of Physics and Astronomy, University College London, Gower Street, London WC1E 6BT, UK}

\date{\today}

\begin{abstract}
Non-resonant X-ray magnetic scattering has been used to study the
magnetic structure of multiferroic TbMnO$_3$ in its ferroelectric
phase. Circularly polarized X-rays were combined with a full
polarization analysis of the scattered beam to reveal important new
information on the magnetic structure of this canonical
multiferroic. An applied electric field is shown to create a
magnetic nearly mono-domain state in which the cylcoidal order on
the Mn sublattice rotates either clockwise or counter-clockwise
depending on the sign of the field. It is demonstrated how this
technique provides sensitivity to the absolute sense of rotation of
the Mn moments, and to components of the ordering on the Tb
sublattice and phase shifts that earlier neutron diffraction
experiments could not resolve.

\end{abstract}



\maketitle

Multiferroic materials exhibit unusual physical properties as a
result of coupling between the various forms of spontaneous ferroic order they display
\cite{Fiebig}. Recently,
considerable interest has been generated following the discovery of a large magneto-electric coupling
in TbMnO$_3$ \cite{KimuraTMO}, which was subsequently shown to be due to the establishment of non-collinear
antiferromagnetic order driving the formation of a ferroelectric state (FE)
\cite{kenzelmann}. A central theme of ensuing studies in TbMnO$_3$ and other systems
has been to  explore the electric field control of magnetism, and correspondingly the magnetic field control of
FE \cite{eerenstein2006,Cheong2007}.
Of paramount importance in
the development of any theory of the magneto-electric coupling in
this class of materials, is a complete and accurate microscopic
description of the magnetic structures they display.

Historically the utility of circularly polarized X-rays is well
established in dichroism experiments on ferromagnets
\cite{Erskine75,vanderLaan86,Schutz87}. Less attention has been paid
to the case of antiferromagnets, with even fewer examples of their
use in diffraction from non-collinear systems such as helices and
cycloids \cite{Sutter,Durr1999}. Here we report on a new application
of circularly polarized X-rays in combination with full polarimetry
of the scattered beam to the refinement of magnetic structures in
magneto-electric multiferroics. Besides the classical advantages
provided by X-ray magnetic diffraction (discrimination between spin and orbital contribution, in non
resonant scattering) \cite{Paolasini08}, we demonstrate how this
technique provides  unique insight into the formation of cycloidal
domains in TbMnO$_3$. A key feature of these experiments was the control of the
population of magnetic domains by an applied electric field.

%%%%%%%%%%%%%%%%%%%%%%%%%%%%%%%%%%%%%%%%%%%%%%%%%%
\begin{figure}
    \begin{center}
        \includegraphics[width=.9\linewidth,bb=50 370 535 760,clip]{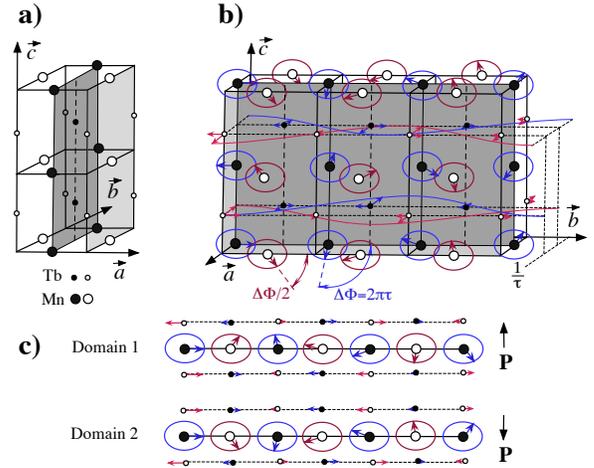}
    \end{center}
    \vspace{-10pt}
\caption{a) Crystallographic and b) magnetic structures of
TbMnO$_3$. The incommensurate magnetic structure propagates along
the $\mathbf b$ direction with periodicity $1/\tau~b_0 \approx 3.55
~b_0$ on both the Mn and Tb sublattices. Both the transverse and the
previously undetermined longitudinal components of the Tb moments
are shown. c) Projection in the $\mathbf{b-c}$ plane of the two
cycloidal magnetic domains, also showing the newly determined
longitudinal Tb moment component. Domain 1 (2) is favoured by
electric field $\mathbf E>0$ ($\mathbf E<0$) (see
Fig.~\ref{fig:annealingresults}). \label{fig:MagnStruct}}
\end{figure}
%%%%%%%%%%%%%%%%%%%%%%%%%%%%%%%%%%%%%%%%%%%%%%%%

Bulk measurements first established that TbMnO$_3$ undergoes a
transition below $T$=27~K to a multiferroic state that is both
magnetic and FE, and demonstrated how a magnetic field could be used
to control ferroelectricity \cite{KimuraTMO}. A major breakthrough
in understanding this effect was provided by a neutron diffraction
study which concluded that at $T$=27~K the Mn $3d$ magnetic moments
undergo a transition from a collinear to a non-collinear, cycloidal
phase, described by an incommensurate wavevector ${\bf
k}_m=\tau\mathbf{b}^*$, which removes a centre of inversion so as to
form a FE state \cite{kenzelmann}.  For this class of multiferroic,
the FE moment $\mathbf P$ may be conveniently written as $\mathbf P
\propto \mathbf k_m \times \mathbf C$, where $\mathbf C$
=$\sum_i\,\mathbf S_i \times \mathbf S_{i+1}$ characterises the
magnetic structure adopted by the spins $\mathbf S_i$
\cite{Katsura2005,Mostovoy2006,Sergienko2006}. In the case of
TbMnO$_3$ the Mn magnetic moments in the cycloidal phase rotate in
the $\bf b - \bf c$ plane (Fig.\ \ref{fig:MagnStruct}), and with
${\bf k}_m$ parallel to the $\bf b$ axis, $\bf P$ is either parallel
or anti-parallel to the $\bf c$ axis depending on the sign of $\bf
C$. The same study  \cite{kenzelmann} proposed that the Tb $4f$
moments are also sinusoidally modulated at ${\bf k}_m$, but are
transversely polarized along the $\bf a$ axis. Polarized neutron
diffraction has recently been applied to multiferroics
\cite{Radaelli2008,lee}, including TbMnO$_3$ \cite{yamasaki}, where
it was found that the cycloid could be switched from propagating
clockwise or anticlockwise, depending on the direction of the
applied electric field \cite{yamasaki}. Our non-resonant X-ray
magnetic scattering (NRXMS) study of TbMnO$_3$ establishes the
utility of X-rays for studying the electric field control of
magnetism in multiferroics. Although NRXMS experiments are in many
ways much more demanding than neutron diffraction, due to the
extreme weakness ($\sim 10^{-5}-10^{-7}$) of the NRXMS relative to
the charge scattering, they are nontheless of considerable value due
to the complementary nature of the information they provide. In the
case of TbMnO$_3$ this has allowed us to reveal the ordering of a
$b$ component on the Tb sublattice and to determine various phase
relationships that were hitherto unknown.

%%%%%%%%%%%%%%%%%%%%%%%%%%%%%%%%%%%%%%%%%%%%%%%%%%
\begin{figure}
    \begin{center}
    \includegraphics[width=.9\linewidth,bb=55 390 490 660,clip]{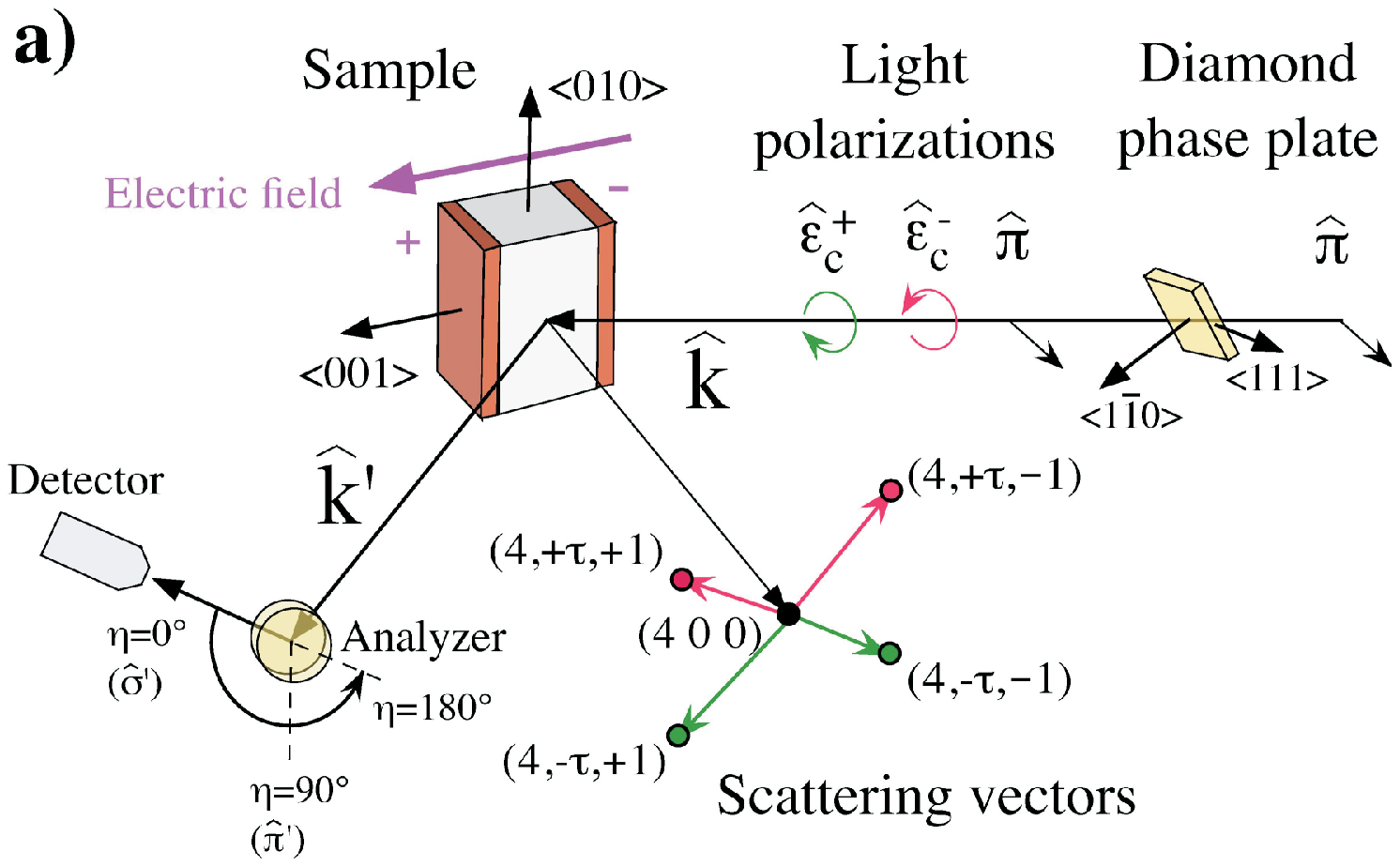}
    \includegraphics[width=.9\linewidth,bb=17 370 569
    727,clip]{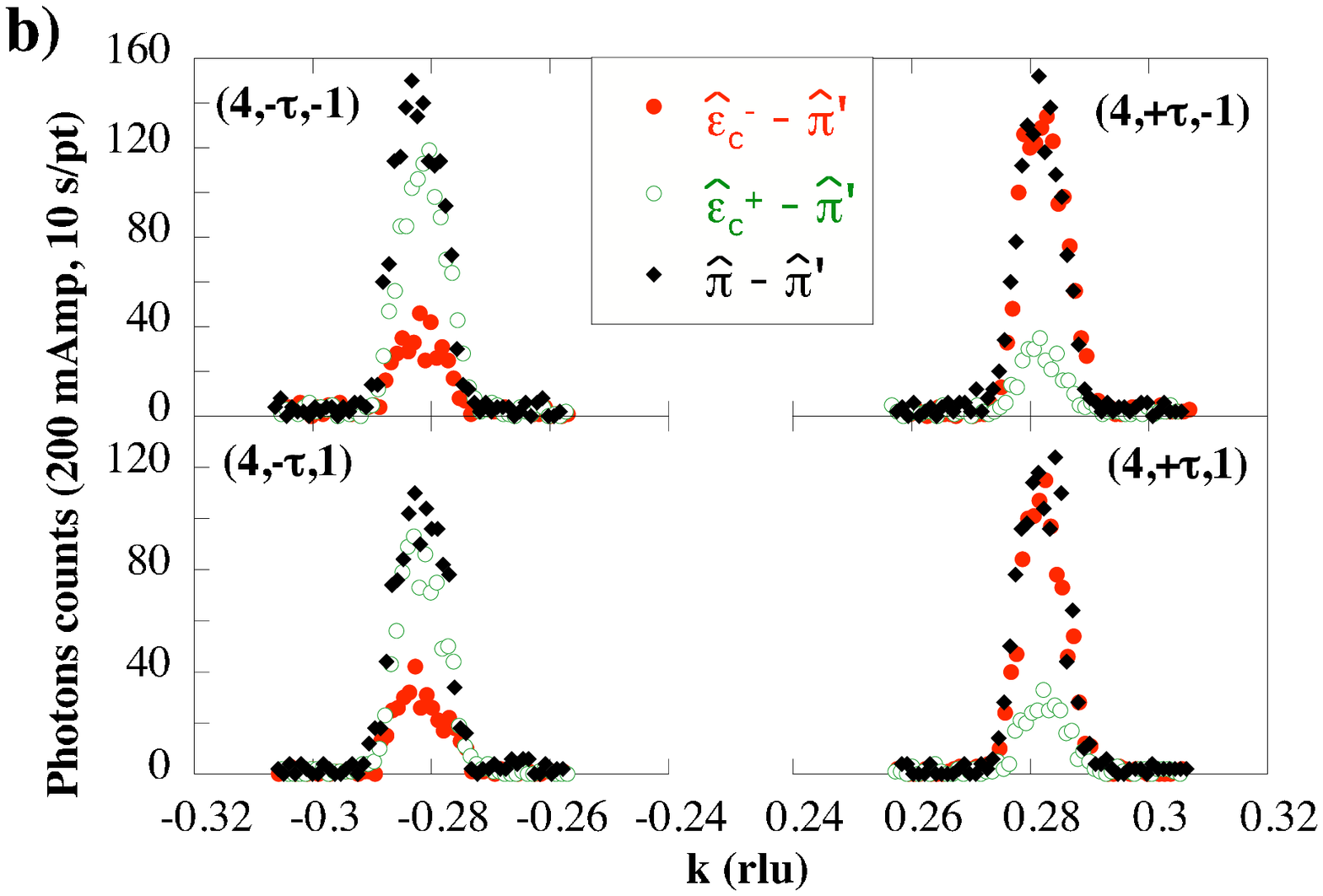}
    \end{center}
    \vspace{-10pt}
\caption{ (a) Schematic of experimental scattering geometry used to
determine the polarization dependence of A-type magnetic satellite
reflections ($4, \pm\tau, \pm1)$ in TbMnO$_3$. (b) Incident
polarization dependence of these reflections in the FE cycloidal
phase at 15~K, obtained after annealing the sample with
$\mathbf{E}<0$. The polarization of the scattered photons was
analyzed in the $\hat{\pi}'$ channel
($\eta=90^{\circ}$).\label{fig:geomandk}}
\end{figure}
%%%%%%%%%%%%%%%%%%%%%%%%%%%%%%%%%%%%%%%%%%%%%%%%

TbMnO$_{3}$ adopts the $Pbnm$ space group at room
temperature with
$a_{0}=5.302$ \AA,  $b_{0}=5.856$ \AA ~and $c_{0}=7.400$ \AA ~\cite{Blasco}. We selected a single crystal
grown by the floating zone method, as used in previous
resonant X-ray scattering experiments \cite{mannix}. The sample had dimensions $2\times
2\times 0.8$ mm$^{3}$, and a poling electric field $\mathbf E$ of about 1~kV/mm
was applied  during the cooling procedure inside a $^4$He evaporation
cryostat. Two copper electrodes were
glued by high conductive silver paste on the sample $c$-surfaces,
and the electric field was removed during the X-ray experiments.

X-ray magnetic diffraction experiments were performed at the ID20
beamline \cite{Paolasini07} (ESRF, Grenoble, France) using a
monochromatic beam at a wavelength of $\lambda$ =1.66 \AA, so as to
excite non-resonant processes, which allows the spin and orbital
magnetizations to be determined independently \cite{Blume}. (Neutron
scattering, by contrast, is sensitive to their sum.) Linearlly
polarized X-rays were converted into left-circularly polarized (LCP)
or right-circularly polarized (RCP) photons using a diamond
quarter-wave plate (circular polarization typically $99\%$), or
allowed to pass unchanged through the phase plate, see
Fig.~\ref{fig:geomandk}(a). For a cycloidal magnet, the incoming
circular polarization finds its handedness naturally coupled to the
sense of rotation of the magnet moments, allowing the population of
domains to be determined by observing differences in the intensity
of magnetic diffraction peaks under illumination with LCP or RCP
X-rays. An Au (222) crystal analyzer system mounted on the detector
arm of a six-circle diffractometer was used to characterise the
polarization of the scattered photons. The polarization is
represented by the Stokes vector {\bf P}=($P_1$,$P_2$,$P_3$)
\cite{deBergevin,Blume}, whose linear components  $P_1$,  $P_2$ were
determined by fitting the dependence of the scattered X-ray
intensity $I(\eta)=I_0\left(1+P_1 \cos 2\eta+P_2 \sin 2\eta\right)$
on the angle $\eta$ of the polarization analyzer about the
wavevector ${\bf k}'$. The Stokes parameters of the incident beam
were carefully checked before and after every set of scans by using
the same polarization analyzer, to control the incident circular
photon polarization, which is very sensitive to the beam position
stability.

The initial objective of our study was to establish whether or not our
experimental setup provided sufficient sensitivity to any imbalance
in the population of magnetic domains produced by the applied
$\mathbf E$ field. Figure~\ref{fig:geomandk}(b) shows the
polarization dependence of the scattering from the four
A-type \cite{Quezel} $(4,\pm\tau,\pm1)$ magnetic satellite
reflections $-$ each satellite forming one arm of the so-called star
of wavevectors around $(4,0,0)$, see Fig.~\ref{fig:geomandk}(a). The
measurements were performed at 15~K in the FE phase, after cooling
the sample with a poling voltage of $-$700~V. Data were collected
with linear ($\hat\pi$, black diamonds), RCP
($\hat\epsilon_c^+$, open green circles), and LCP
($\hat\epsilon_c^-$, red filled circles) polarized X-rays, while the
scattered beam was analyzed in the $\hat\pi'$ channel
($\eta=90^\circ$). Inspection of Fig.~\ref{fig:geomandk}(b) reveals
that the intensities of the magnetic satellites are very similar when
incident linear polarization $\hat{\pi}$ is selected. For incident
circular polarizations this is not the case: instead the intensities
display complementary behaviour, depending on the sign of $\tau$. In
the case of equi-populated cycloidal domains the intensities
associated with $\hat\epsilon_c^+$ and $\hat\epsilon_c^-$ should be
similar, whereas for a single cycloidal domain a large difference is
expected, as is indeed observed to be the case.

\begin{figure}
    \begin{center}
        \includegraphics[width=.99\linewidth,bb=44 240 548 755,clip]{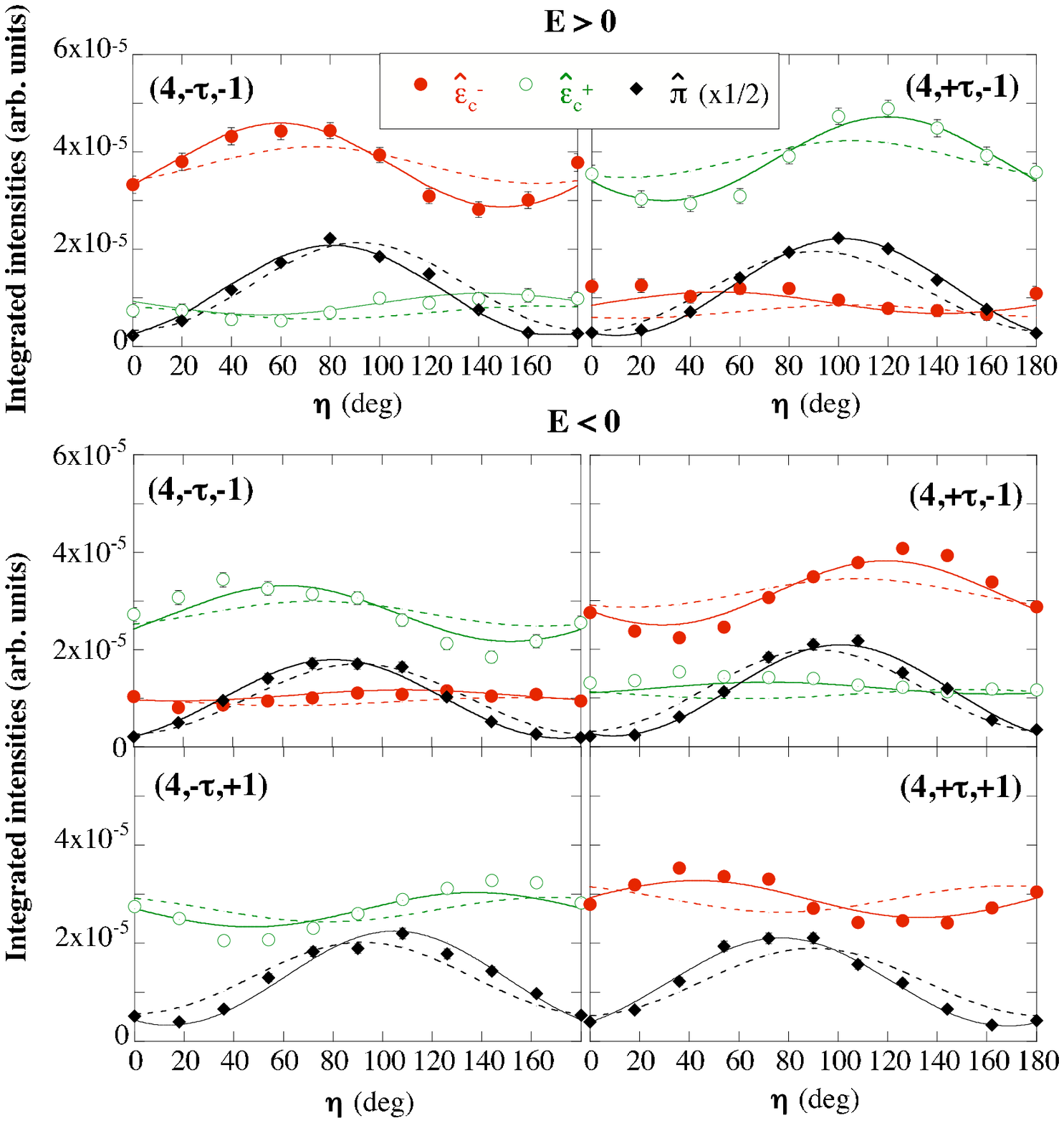}
    \end{center}
    \vspace{-10pt}
\caption{Variation with analyzer rotation angle $\eta$ of the X-ray
magnetic scattering in TbMnO$_3$ at 15~K, for field cooling with
either $\mathbf{E}>0$ (upper panel) or $\mathbf{E}<0$ (lower panel).
The dashed lines represent the  model described in Ref.
\cite{kenzelmann}, whereas the continuous lines are our model
calculations, as described in the text.\label{fig:annealingresults}}
\end{figure}

The standard approach to obtain a microscopic understanding of the
field-induced magnetic domain state would be to collect intensity
data for a set of satellites ($h$,$k\pm\tau$,$l$) with $(hkl)$
taking on as wide a range of values as possible. The disadvantages
of this method are that large corrections must be applied for
effects such as absorption, {\it etc.}, and that our specific
scattering geometry restricts the accessible $(hkl)$ values. Instead
we undertook a detailed analysis of the polarization of the
scattered beam by carefully measuring $I(\eta)$ for each of the
satellites shown in  Fig.~\ref{fig:geomandk}(a), extracting the
Stokes parameters. Figure~\ref{fig:annealingresults} summarises our
data for field-coolings performed with either $\mathbf E>0$ or
$\mathbf E<0$. The data were obtained by rocking the crystal
analyzer at different $\eta$ angles, and are normalized by  monitor
and corrected by subtraction of the background, measured off peak.

%%%%%%%%%%%%%%%%%%%%%%%%%%%%%%%%%%%%%%%%%%%%%%%%%
%%%%%%%%%%%%%%%%%%%%%%%%%%%%%%%%%%%%%%%%%%%%%%%

Turning  first to consider the case of incident linear ($\hat\pi$)
polarization shown in Fig.\ \ref{fig:annealingresults}, it is clear
that for a particular choice of the direction of $\mathbf E$, only
small differences in  $I(\eta)$ are observed for the various
satellites, and that reversing the direction of $\mathbf E$ does not
appear to have any effect. Clear differences in $I(\eta)$ of the
various satellites are observed for incident circular polarization.
In this case reversing the direction of $\mathbf E$ has a profound
effect on the observed intensities. For a given field direction,
changing the sign of $\tau$ at fixed $l$ leads to a switching of the
handedness of incident X-rays that produces the maximum intensity,
whereas the handedness associated with the maximum satellite
intensity is invariant with respect to changes in the sign of $l$.
Of considerable significance is the fact that reversing the sign of
$\mathbf E$ leads to a switching of the dominant handedness. Also
apparent are clear mirror symmetries between LCP and RCP X-rays for
the same reflection, and between $\pm l$ satellites for the same
incident polarization state. Thus at a qualitative level the data
displayed in Fig.~\ref{fig:annealingresults} reveal an imbalance in
the population of the two possible magnetic domains created by the
applied electric field, an imbalance that can be reversed by
switching the direction of the field.

To explore the potential richness of information encoded in the data
shown in Fig.~\ref{fig:annealingresults} we performed extensive
modelling of the magnetic scattering. For the initial model we used
the structure due to Blasco {\it et al.} \cite{Blasco}, as shown in
Fig.~\ref{fig:MagnStruct}(a), and the magnetic structure proposed by
Kenzelmann {\it et al.} \cite{kenzelmann}, as shown in
Fig.~\ref{fig:MagnStruct}(b). In Fig.~\ref{fig:annealingresults} the
results of our calculations of NRXMS \cite{Blume} based on this model are represented
by dashed lines. For the calculations the exact scattering geometry
of the six-circle setup was taken into account.
The agreement
between the model and the data is clearly unsatisfactory. In
particular, the data show a non zero value of $P_2$
(a linear polarization oblique with respect to directions $\hat{\sigma}'$, $\hat{\pi}'$),
that is not predicted by the model.

A vastly improved description of the data was achieved following the
realisation that, in contrast to the earlier neutron diffraction
study, our experiment yielded sensitivity to the $b$ component of
the ordered moment on the Tb sites, while being largely insensitive
to the $a$ component. From the fits we are   able to fix the phase
angle between the magnetic modulations of the  Tb and the Mn atoms.
Consider one Mn atom, and the subsequent Tb atom moving along  $\pm
\mathbf{c}$ (sign determined accordingly to the direction of the
electric polarization), then the phase angle between their magnetic
modulations, evaluated at the same coordinate in space, is found to
be 1.0 $\pm$ 0.1 $\pi$ (with reference  to the $b$ components).
The phase shift between the two orbits of Tb atoms (at $z = 1/4$ and $z = 3/4$ of the unit cell) is
determined to be 1.0 $\pm$ 0.2 $\pi$.

The model results in the following NRXMS
amplitude for an A-type peak of the form (4 $\pm \tau$ $\pm$1):
\begin{equation*}  \label{ScattAmpl}
\begin{split}
&f({\bf K})  \ \propto \  ( S_b^{Mn}({\bf K}) \hat{{\bf b}} \
- \gamma ~ \alpha \ i\  S_c^{Mn}({\bf K})  \hat{{\bf c}}  ) \cdot {\bf B}\\
&-\ \gamma ~  \beta \ i \ ( S_b^{Tb}({\bf K}) \hat{{\bf b}}  \cdot
{\bf B}    +  \frac{1}{2}  L_b^{Tb}({\bf K})  \hat{{\bf b}}  \cdot
{\bf A}'' ) \cos(8\pi \Delta^{Tb}_a)
\end{split}
\end{equation*}
where  the value of $\alpha=\pm 1$ selects the sign of $\tau$,
$\beta$ the sign of  $l = \pm 1$ and $\gamma=\pm 1$ identifies the
two cycloidal domains 1 and 2, respectively
(Fig.~\ref{fig:MagnStruct}(c)). The vectors ${\bf A''}$ and ${\bf
B}$ contain the dependence on the polarization of the incident and
diffracted X-rays \cite{Blume}. The spin $S_i$({\bf K}) and the
orbital $L_i$({\bf K}) magnetic components contain the form factors
for Mn$^{3+}$ and Tb$^{3+}$ at a given scattering vector
$\mathbf{K}$. We suppose also that the orbital contribution of Mn is
quenched, and that of Tb equal to its spin moment
($\mathrm{L}^{Tb}/\mathrm{S}^{Tb}=1$) as follows from Hund's rules
for a $^7F_6$ electronic configuration. Finally $\Delta^{Tb}_a$
describes the fractional atomic coordinate along the $\mathbf{a}$
axis of the Wyckoff position (4c), for the Tb atoms \cite{Blasco}.

A simplified expression for the NRXMS  amplitude in the
$\sigma^\prime$ and $\pi^\prime$ channels, in which we set the
scattering angle $2\theta=90^\circ$, $\Delta_a^{Tb}=0$ and place the
$\mathbf{b}$ axis perpendicular to the scattering plane, illustrates
the sensitivity in our experiment to the different components of the
Tb and Mn magnetic moments, and the dependence of the scattered
intensity on the handedness of the X-rays, etc:
\begin{eqnarray*}
f_{\hat{\sigma} '}&\propto& S_b^{Mn} + \epsilon ~ \alpha ~ \gamma ~ {S'}_c^{Mn}  - i ~ \beta ~ \gamma ~ S_b^{Tb} \\
f_{\hat{\pi} '}&\propto& (\epsilon ~ \beta ~ \gamma) (S_b^{Tb} +
L_b^{Tb}) + i ~ (\epsilon ~ S_b^{Mn} + \alpha ~ \gamma ~ S_c^{Mn}).
\end{eqnarray*}
Here $\epsilon$ selects the handedness of the incident X-rays, and
${S^{(\prime)}}_c^{Mn}$, {\it etc.}, are the projected magnetization
densities on the incident (scattered) wavevector. The dominant Mn
terms are real in $f_{\hat{\sigma}'}$ and imaginary in
$f_{\hat{\pi}'}$, giving a scattered beam that is mainly circularly
polarized. They are large (small) if
$\epsilon\alpha\gamma=+1\,(-1)$. A non zero value of $P_2$  arises
when the pair $f_{\hat{\sigma}'} \pm f_{\hat{\pi}'}$ have different
moduli. This depends on $S_b^{Tb}$ and  $L_b^{Tb}$. Their sign in
$f_{\hat{\pi}'}$ ($f_{\hat{\sigma}'}$) is to be compared to the sign
of the largest term $S_b^{Mn}$ in $f_{\hat{\sigma}'}$
($f_{\hat{\pi}'}$). Then it is seen that $S_b^{Tb}$ adds with
opposite signs in the real and imaginary parts, so that its effect
in $P_2$ nearly cancels. Instead $L_b^{Tb}$ produces a net $P_2$
with the sign $+\epsilon \beta\gamma$.

The full expression  \ref{ScattAmpl} for $f({\bf K})$ was fitted to
the data with the outcome represented by the solid lines in
Fig.~\ref{fig:annealingresults}. It is clear that inclusion of a $b$
component of the Tb moment leads to an excellent description of the
data, with the best fit obtained with  (${\mathrm L}_b^{Tb}$ +
2${\mathrm S}_b^{Tb}$) equal to 1.0 $\pm$ 0.3 $\mu_B$. Most
importantly, our refined model of the magnetic structure allows us
to obtain an accurate description of the domain state in our sample
as a function of applied electric field, including the absolute
sense of rotation of the magnetic moments. We conclude that cooling
in a positive electric field led to a population of 96(3)\% of the
cycloidal domain in which the transverse spiral of the Mn atoms is
counterclockwise, when moving along $+\mathbf{b}$ and looking from
$+\mathbf{a}$ (domain 1 in Fig.~\ref{fig:MagnStruct}(c)), while
field cooling in a  negative electric field produced a population of
83(2)\% for the domain of opposite sense of rotation (domain 2 in
Fig.~\ref{fig:MagnStruct}(c)).

Our results establish the benefits of performing NRXMS experiments
when circularly polarized X-rays are combined with full polarization
analysis of the scattered beam. Applied to cycloidal systems this
technique is capable of providing a quantitative, microscopic
description of the cycloidal domain state. For the specific case of
multiferroic TbMnO$_3$, with two different types of magnetic ion,
making it by any standards a challenging test case, we have shown
surprisingly that this approach allows us to make important
refinements to the magnetic structure obtained from neutron
diffraction. This technique could readily be applied to other
multiferroics, particularly those such as BiFeO$_3$
\cite{Teague,Sosnowska} for which open questions remain concerning
the magnetic structure of bulk samples and its modification in thin
films \cite{Wang03,Ederer,Zhao06,lee}.

\begin{acknowledgments} We wish to acknowledge Andrea Fondacaro for technical support at ID20.
\end{acknowledgments}

\bibliography{fabriziprl}

\end{document}